# First principle study on electronic structure of ferroelectric $PbFe_{0.5}Nb_{0.5}O_3$

Yuan Xu WANG, C L WANG, M L ZHAO and J L ZHANG
*School of Physics and Microelectronics,*
*Shandong University, Jinan 250100, P R China*
Email: wangyx@sdu.edu.cn

**Abstract**

The full potential linearized augmented plane wave (FLAPW) method was used to study the crystal structure and electronic structure properties of $PbFe_{0.5}Nb_{0.5}O_3$ (PFN). The optimized crystal structure, density of states, band structure and electron density distribution have been obtained to understand the ferroelectric behavior of PFN. From the density of states analysis, it is shown that there is a hybridization of Fe d - O p and Nb d - O p in ferroelectric PFN. This is consistent with the calculation of electronic band structure. This hybridization is responsible for the tendency to its ferroelectricity.

<u>*Key words*</u>: A. Ferroelectrics; D. Electronic band structure

## **Introduction**

Perovskite $ABO_3$ materials have been extensively studied because of their technical importance and the fundamental interest in the physics of their phase transitions. Most perovskite ferroelectrics that have been widely used are solid solutions(such as $PbZr_{1-x}Ti_xO_3$ i.e. PZT) and complex compounds(two or more kinds of atoms occupy the same crystallographic positions, such as $PbMg_{1/3}Nb_{2/3}O_3$). First principle calculations on these complicated systems are very interesting. Lead Iron Niobate PbFe$_{0.5}$Nb$_{0.5}$O$_3$ (PFN) is a complex compound ferroelectric with Fe and Nb occupy the B-site in the ABO$_3$ perovskite structure. Its ferroelectricity at room temperature were discovered in 1957 by Smolenskii et al. [1]. Since then, many people studied its phase transition and other properties [2-5]. Apart from the bulk ceramic, the ferroelectric, dielectric and leakage current properties show that it is potential material for nonvolatile ferroelectric memory and high dielectric constant DRAM applications [6]. At 376 K, it undergoes a ferroelectric phase transition from cubic phase to tetragonal phase [7]. The occupation of Fe and Nb in the site B was examined by transmission electron microscopy (TEM) [6]. The (1/2, 1/2, 1/2) superlattice reflection plots which indicate the ordering of the occupation of irons in site B could not be found, thus confirming that Fe and Nb randomly distributed on site B. Its fine structure was investigated in our previous paper [8]. To our knowledge, there is little investigation on electronic structure of PFN by first principles calculation. In order to well know the origin of its ferroelectricity, we calculated its electronic structure (density of states, band structure) by FLAPW method within the generalized gradient approximation (GGA).

**Approach**

The calculations presented in this work were performed within the generalized gradient approximation (GGA)[9] to density functional theory, using the FLAPW method (WIEN2k [10]). In order to calculate the electronic structure of PbFe$_{0.5}$Nb$_{0.5}$O$_3$, we construct a supercell for it. In the [001] direction, there are two cells, PbFeO$_3$ and PbNbO$_3$. The system structure parameters were optimized by first principle calculation. In electronic structure calculation, we use our optimized parameters. The atomic sphere radii ($R_i$) 2.0, 1.8, 1.9 and 1.6, were used for Pb, Fe, Nb, and O, respectively. We used 7×7×3 meshes and $RK_{max}$ = 8.0 to calculate the GGA results.

**Results and Discussion**

Figure1 shows our calculated total energy of PFN as a function of cell volumes. Our theoretical lattice parameters is a = 3.9885 Å and experimental lattice parameter is a= 4.010 (cubic phase) and a = 4.007 Å, c = 4.013 Å [7]. Our theoretical lattice parameters are close to experimental results. Then we optimized its atomic coordinates.

Table 1  Theoretical atomic coordinates

| Element | x | y | z |
|---|---|---|---|
| O | 0 | 0.5 | 0.25 |
| O | 0 | 0.5 | 0.752189 |
| O | 0.5 | 0 | 0.25 |
| O | 0.5 | 0 | 0.748394 |
| O | 0.5 | 0.5 | 0 |
| O | 0.5 | 0.5 | 0.5 |
| Fe | 0.5 | 0.5 | 0.745723 |
| Nb | 0.5 | 0.5 | 0.246352 |
| Pb | 0 | 0 | -0.039843 |
| Pb | 0 | 0 | 0.531113 |

Table 1 shows our theoretical optimized atomic coordinates. In order to understand the role of Fe and Nb in ferroelectric PFN, we calculated its electronic structure. Figure 2 shows the density of states of ferroelectric PFN. In Figure 2, the DOS shape of atom Nb d and Fe d are very similar. There are two sharp peaks on about 0 eV and 2 eV and a broad peak in valence energy range. This similar character shows in ferroelectric phase, Fe d and Nb d electrons at the same energy range increase greatly. Another, in the energy range from –6 to –2 eV, B (Fe and Nb) d and O p has a broad DOS peak and the peaks of three atoms are similar. It means there is a hybridization between B (Fe and Nb) d and O p and this hybridization weakens the short range repulsions and stabilizes the ferroelectricity. This strong hybridization also implies the interaction between atom B and O is highly covalent. Another, the sharp peaks around of Fe d and Nb d indicate the ferroelectric PFN is more electric conductive than normal ferroelectrics.

Figure 3 shows the band structure of ferroelectric PFN. From -6 to -2 eV, there are many bands. Comparing with Figure 2 in the same energy range, it can be seen that these bands are mainly from O p and some of B d electrons. These bands overlap each other. It means that there is a hybridization between the B d and O p. The same conclusion has been obtained from the partial DOS analysis of B d and O p. Around and above 0 eV, some conduction bands exist. Comparing with Figure 2, it can assessed that these bands mainly consist of B d and a little of O p. The existence of these bands indicate that ferroelectric PFN is more electric conductive than normal ferroelectrics and this is consistent with the experimental result [11]. In ref [11], authors show their experimental result of electric conductivity of PFN. Another, there are two bands around -9 eV. The two bands should come from Pb 6s. The large dispersion of these two bands implies its free character. Like ferroelectric $PbTiO_3$, Pb 6s do lie very close to the oxygen bands, which lead to the large Pb-O hybridization. So atom Pb in the ferroelectric PFN plays a similar role to that in pure $PbTiO_3$.

Figure 4 shows the electron density distribution of ferroelectric PFN in (100) plane. In Figure 4, we can see that the interaction between atom Fe and O, Nb and O is very strong. It means the strong hybridization between B(Ta and Nb) d and O p.

## Conclusion

In summary, we have performed FLAPW calculations to investigate the ferroelectricity of $PbFe_{0.5}Nb_{0.5}O_3$. The ground state structure and electronic properties are obtained. From the calculated DOS and electronic band structure, it is shown that there is a strong hybridization between B (Fe and Nb) d and O p electrons. This hybridization is important to the ferroelectric stability of PFN. In addition, from the band structure analysis, the atom Pb has a significant effect on ferroelectricity of PFN and thus in ferroelectric PFN, atoms Pb play a important role as they do in pure $PbTiO_3$.

## Acknowledgments

This work was supported by a grant for State Key Program for Basic Research of China and Younger Natural Scientific Foundation of Shandong University.

FIGURE CAPTIONS

Figure 1. Total energy of ferroelectric PFN

Figure 2. The DOS of Fe d, Nb d and O p in ferroelectric phase

Figure 3. The band structure of ferroelectric PFN

Figure 4. The electron density distribution of ferroelectric PFN in the (100) plane

Figure 1

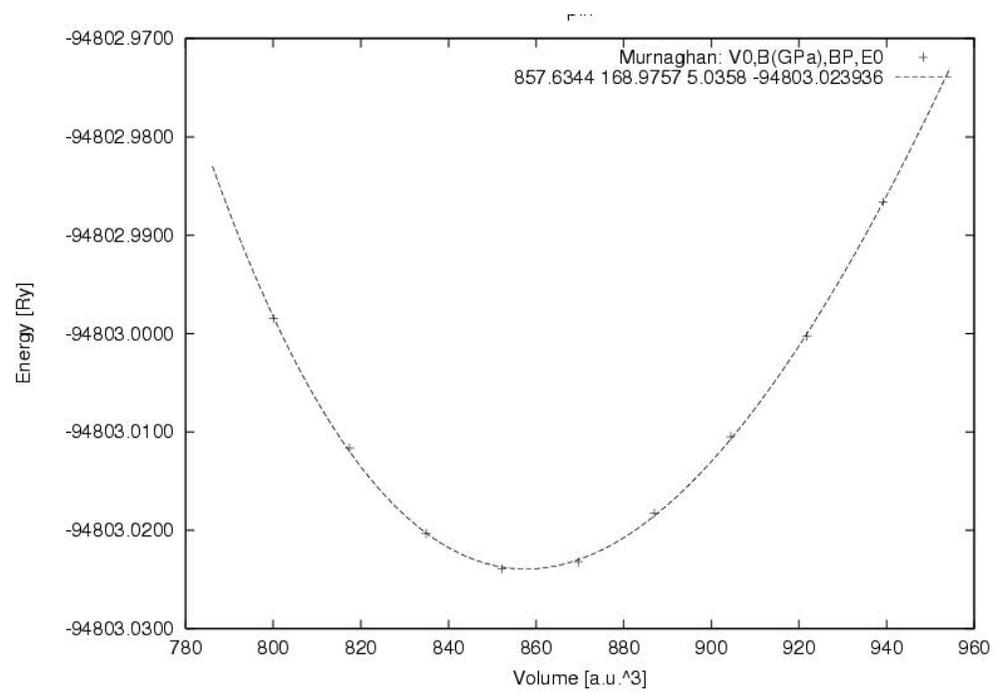

Figure 2

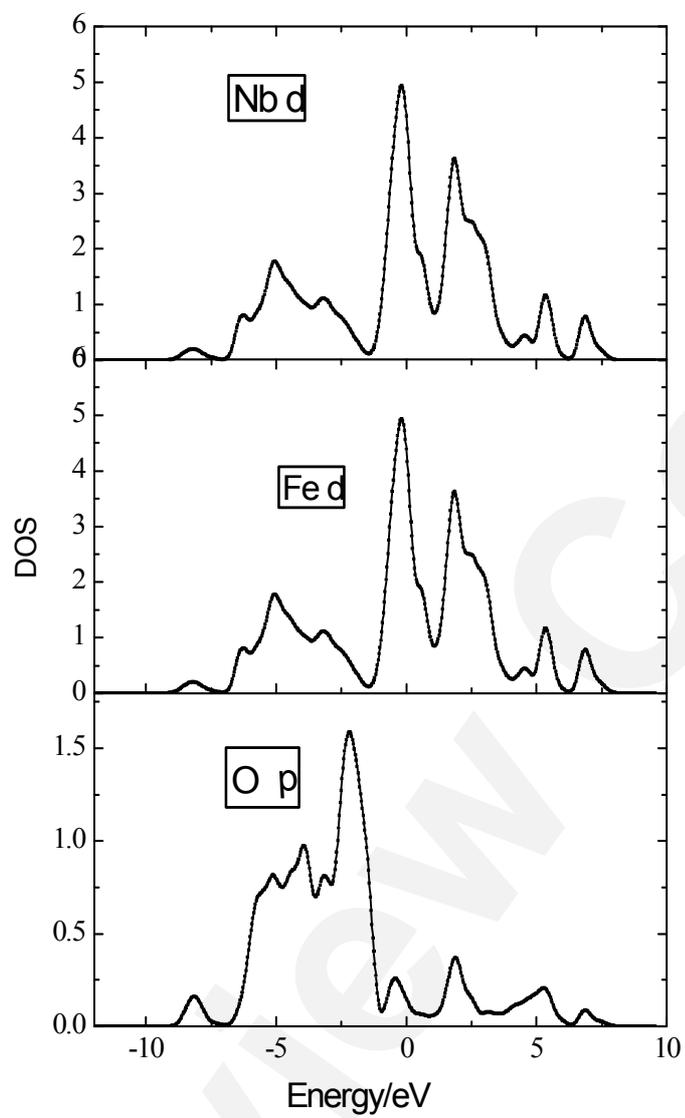

Figure 3

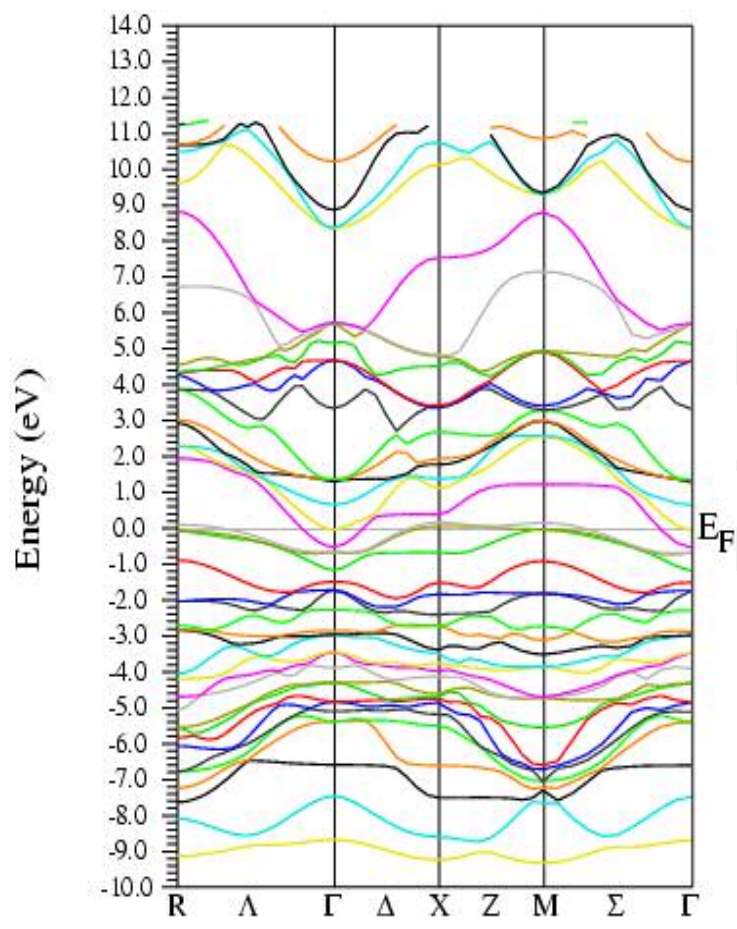

Figure 4

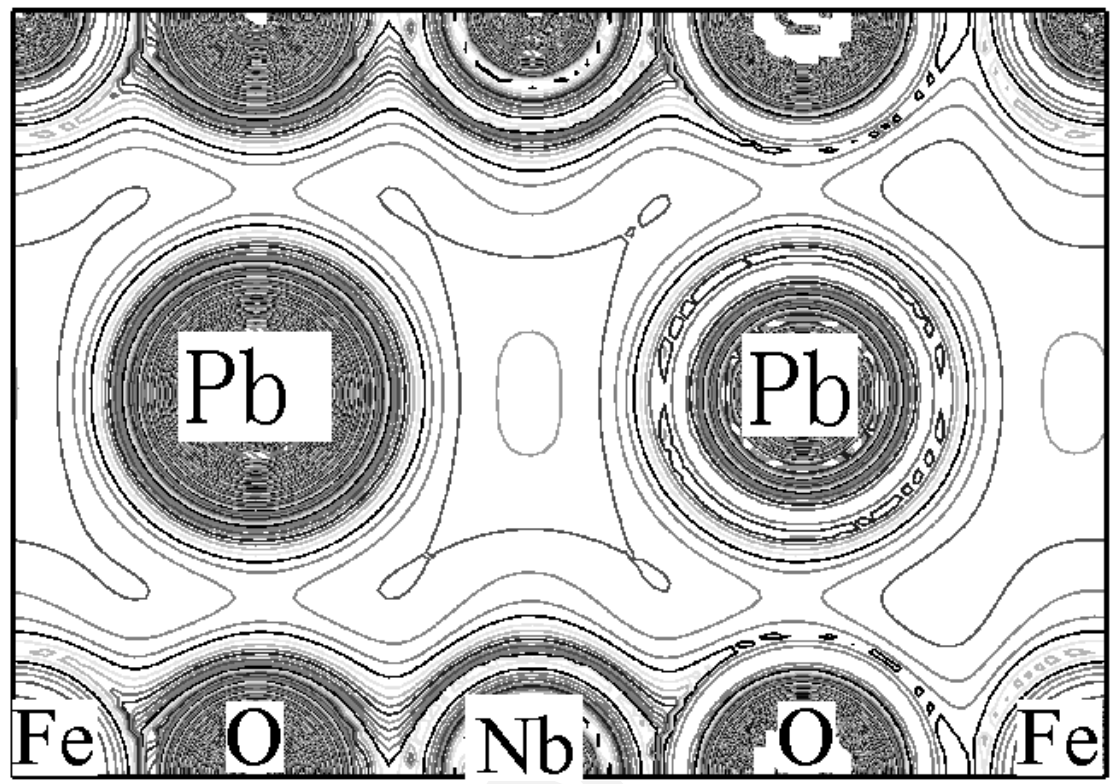

10